\documentclass[preprint,showpacs,tightenlines,amsmath,amssymb]{revtex4}
\usepackage{graphicx}
\begin{document}
\date{\today}
\title{Atomic Parity Non-Conservation, \\
       Neutron Radii, and Effective Field Theories of Nuclei}
\author{Tapas Sil, M. Centelles, and X. Vi\~nas}
\affiliation{Departament d'Estructura i Constituents de la Mat\`eria,
     Facultat de F\'{\i}sica,
\\   Universitat de Barcelona,
     Diagonal {\sl 647}, {\sl 08028} Barcelona, Spain}
\author{J. Piekarewicz}
\affiliation{Department of Physics, Florida State University,
Tallahassee, Florida {\sl 32306}}
% >>>>>>>>>>>>>>>>>>>>>>>>>>>>>>>>>>>>>>>>>>>>>>>>>>>>>>>>>>>>>>>>>>>>

\begin{abstract}
Accurately calibrated effective field theories are used to compute
atomic parity non-conserving (APNC) observables. Although accurately
calibrated, these effective field theories predict a large spread in
the neutron skin of heavy nuclei. While the neutron skin is strongly
correlated to a large number of physical observables, in this
contribution we focus on its impact on {\it new physics} through APNC
observables. The addition of an isoscalar-isovector coupling constant
to the effective Lagrangian generates a wide range of values for the
neutron skin of heavy nuclei without compromising the success of the
model in reproducing well constrained nuclear observables. Earlier
studies have suggested that the use of isotopic ratios of APNC
observables may eliminate their sensitivity to atomic structure. This
leaves nuclear structure uncertainties as the main impediment for
identifying physics beyond the standard model. We establish that
uncertainties in the neutron skin of heavy nuclei are at present too
large to measure isotopic ratios to better than the 0.1\%
accuracy required to test the standard model. However, we argue that
such uncertainties will be significantly reduced by the upcoming
measurement of the neutron radius in ${}^{208}$Pb at the Jefferson
Laboratory. 
\end{abstract}
\maketitle
% >>>>>>>>>>>>>>>>>>>>>>>>>>>>>>>>>>>>>>>>>>>>>>>>>>>>>>>>>>>>>>>>>>>>
% >>>>>>>>>>>>>>>>>>>>>>>>>>>>>>>>>>>>>>>>>>>>>>>>>>>>>>>>>>>>>>>>>>>>

\section{Introduction}

The quest for discovering fundamental characteristics of atomic
nuclei, such as their sizes and shapes, is as old as Nuclear Physics
itself. Although for the most part this challenge has been met with
enormous success, precise knowledge of certain fundamental
nuclear-structure properties---among them the spatial distribution of
neutrons---is still lacking. While the original motivation for
measuring the neutron density of atomic nuclei was deeply rooted in
nuclear structure, the quest has been recently revived in response to
the widespread impact that such a measurement will have over seemingly
unrelated areas of physics. Even though the present study concentrates
exclusively on atomic parity non-conservation (APNC)
\cite{Bouchiat97,Ginges04}, a brief review of areas that could benefit
from an accurate measurement of the neutron density is also presented.

Unquestionably, the accurate determination of the neutron
density---and in particular its root-mean-square (rms)
radius---remains a top priority in nuclear-structure physics. This
inadequate state of affairs is in stark contrast to our present
knowledge of proton densities. Indeed, the proton distribution in
nuclei has already been mapped with exquisite precision, often much
better than 1\% \cite{ang1}, due to the availability of
state-of-the-art electron-scattering facilities all over the world. A
particularly illustrative example is ${}^{208}$Pb, a nucleus whose
charge radius is presently known with an extremely high accuracy
($r_{\rm ch}=5.5010 \pm 0.0009$~fm~\cite{ang1}) that exceeds at least
by two orders of magnitude that of its neutron radius. Clearly, an
accurate determination of neutron radii is a pressing issue. Further,
without such knowledge for stable nuclei, the prospects of real
progress in the future domain of nuclear structure, namely, {\it
exotic nuclei}, may be seriously compromised.

Precision studies of the nucleon-nucleon (NN) force is an area that
has been directly affected by our poor knowledge of neutron radii. At
the most fundamental level, Quantum Chromodynamics predicts a small
flavor violation in the strong force due to the different charges and
masses of the up and down quarks. In turn, this flavor violation
induces a small (of the order of 1\%) breaking in the isospin symmetry
of the NN force. However, to accurately quantify these ``novel''
isospin violations, it is necessary to have all ``conventional''
effects under control. A good example of this interplay between novel
and conventional is provided by the Nolen-Schiffer (or Coulomb energy)
anomaly. The anomaly consists in the residual discrepancy observed in
the ground-state energies of mirror nuclei with a single nucleon added
(or removed) from a closed shell (e.g., ${}^{41}$Sc--${}^{41}$Ca).
However, some of the conventional effects depend sensitively on our
knowledge of neutron radii. Indeed, it has been argued that by using
neutron skins---defined as the difference between the neutron and
proton rms radii---significantly smaller than those predicted by
theory (and apparently justified by experiment) the anomaly
disappears~\cite{shl1}. 

Knowledge of neutron radii could also remarkably improve our
understanding of the equation of state of neutron-rich matter.
Although existing ground-state observables appear to constrain the
symmetry energy of nuclear matter at a neutron density around $\rho_n=
0.10$ fm$^{-3}$ (as a reference value typical in nuclei)
\cite{bro1,furns1}, the density dependence of the symmetry energy is
unknown. Recently, however, it has been found that the neutron skin of
a heavy nucleus, like $^{208}$Pb, calculated with different
non-relativistic and relativistic mean field parametrizations displays
a tight linear relationship with the slope of the equation of state of
neutron matter (or, analogously, with the derivative of the symmetry
energy), evaluated at $\rho_n \sim 0.10$ fm$^{-3}$
\cite{bro1,typ1,cen2,furns1,bald1}. In particular, models with a
``stiffer'' symmetry energy predict larger neutron skins. For
instance, the neutron skin in $^{208}$Pb computed with relativistic
models is larger by typically 0.1--0.2~fm than the value obtained with
Skyrme forces. In turn, the neutron skin of $^{208}$Pb is linearly
correlated with the neutron skin of other relatively heavy nuclei such
as $^{132}$Sn or $^{138}$Ba. Thus, a measurement of the neutron radius
of ${}^{208}$Pb, or indeed of any other heavy nucleus, could fix a
fundamental property of the equation of state.

Fixing the slope of the symmetry energy at $\rho\!=\!0.10~{\rm
fm}^{-3}$ will also impact favorably on astrophysical observables,
particularly on the structure and dynamics of neutron stars. Indeed,
strong correlations between the neutron skin of $^{208}$Pb and various
neutron-star properties have been established~\cite{hor1}. Among
these, a model-independent relation between the neutron skin of
$^{208}$Pb and the transition density from uniform neutron-rich matter
(in the mantle) to a non-uniform phase (in the crust) was observed.
Other neutron-star properties sensitive to the neutron skin of
$^{208}$Pb include the radius of the star, its composition, and its
cooling mechanism \cite{hor2,hor3}. Ultimately, these correlations
emerge as a result of the similar composition of the neutron skin of a
heavy nucleus and the crust of a neutron star, namely, neutron-rich
matter at similar densities.

While the above arguments are already compelling enough to justify the
commission of new experiments to measure the neutron radius of
$^{208}$Pb with unprecedented accuracy~\cite{mic1}, this manuscript
will focus on the crucial role that such a measurement could have on
precision studies of the standard model, vis-\`a-vis, atomic
parity non-conservation. The road to new physics beyond the standard
model in low-energy tests in atoms, which stems from violations of
fundamental symmetries that occur in the weak interaction, passes
through the observation of deviations between the values measured in
the laboratory and the predictions of the standard model. The effects
are inherently small and, traditionally, the analysis of APNC
experiments has been hindered by uncertainties in both atomic and
nuclear-structure theory. However, a fruitful strategy has been
devised to remove the sensitivity to atomic theory. This strategy
consists in measuring {\it ratios} of parity-violating observables
along an isotopic chain~\cite{Dzuba86}. In this manner, uncertainties
in the atomic theory factor out from the ratios. Based on their long
chains of several naturally occurring isotopes, cesium, barium,
dysprosium, and ytterbium appear as ideal
candidates~\cite{Bouchiat97,Ginges04,ben1,xio1,dav1,bud1}. As a
result, nuclear-structure uncertainties---primarily in the form of
neutron radii---remain as the limiting factor in the search for
physics beyond the standard model~\cite{for1,pol2,che1,ram1}.

In an earlier study, a strong correlation between the neutron radius
of $^{208}$Pb and the neutron radii of $^{138}$Ba, $^{158}$Dy, and
$^{176}$Yb was established~\cite{tod1}. In that study the density
dependence of the symmetry energy---and consequently the neutron
radius of $^{208}$Pb---was modified through the addition of a new
coupling constant ($\Lambda_{\rm v}$) to the underlying
Lagrangian~\cite{hor1,hor2}. This new coupling allows one to modify
the density dependence of the symmetry energy without changing the
saturation properties of symmetric nuclear matter. In a heavy nucleus
$\Lambda_{\rm v}$ modifies the rms radius of the neutron density while
leaving the rms radius of the proton density and the total binding
energy practically unchanged \cite{hor1,hor2}. However, the
nuclear-structure model used lacked both deformation and pairing
effects that are important for the nuclei considered in APNC
experiments. Thus, in Ref.~\cite{tod1} the calculations were limited
to the study of a single member of each isotopic chain (the one with a
closed neutron shell). This shortcoming is resolved in the present
contribution. As a result, one is now able to map the neutron radii of
these long isotopic chains as a function of both the neutron number
$N$ and $\Lambda_{\rm v}$. There are previous studies of the isotopic
dependence of the neutron radius in the relativistic mean field (RMF)
approximation~\cite{pan1,vre1} and of its dependence on $\Lambda_{\rm
v}$~\cite{tod1}. Yet the merit of the present work is that
uncertainties in the neutron radius are studied for the first time as
a function of both $N$ and $\Lambda_{\rm v}$ within a unified
``best-fit'' model. In addition, it has been recently pointed out that
the accuracy in the measurements of APNC effects can be improved by
using high-$Z$ atoms, such as francium, where the parity violating
effects are expected to be an order of magnitude larger than in cesium
\cite{spr1}. Thus we also include, in anticipation of future
experiments, the analysis of the nuclear structure corrections to the
weak charge in francium isotopes.

The manuscript has been organized as follows. In Sec.~\ref{Theory} a
short description of the nuclear model is given, with special emphasis
paid to the treatment of deformation and pairing correlations
as well as to the predictions of the model applied to cesium, barium,
dysprosium, and ytterbium. In Sec.~\ref{Results} a brief review of
atomic parity violation is given, with the main goal of addressing
uncertainties in APNC observables emerging from our poor knowledge of
neutron radii; a detailed analysis of such uncertainties within our
model is also presented in the same section. Finally the conclusions
are reserved for Sec.~\ref{Conclusions}. We may summarize the two main
conclusions of our work as follows. First, it appears that even when
using the wealth of existing ground-state observables to constrain our
nuclear-structure models, the uncertainties in the neutron radius of
heavy nuclei---and consequently on APNC observables---may have been
underestimated. Second, we have found, as others have before us, a
tight correlation between the neutron radius of $^{208}$Pb and the
neutron radii of barium, dysprosium, and ytterbium, which is shown
here to hold also in deformed systems and in the presence of pairing.
This correlation suggests that the nuclear structure observables
relevant for APNC will profit positively from the upcoming high
precision measurement of the neutron radius of $^{208}$Pb at the
Jefferson Laboratory~\cite{mic1}.

% >>>>>>>>>>>>>>>>>>>>>>>>>>>>>>>>>>>>>>>>>>>>>>>>>>>>>>>>>>>>>>>>>>>>

\section{The nuclear model}
\label{Theory}

The mean field treatment of effective field theories of hadrons,
generally known as quantum hadrodynamics (QHD), is well established as
a successful approach for describing diverse bulk and single-particle
properties of finite nuclei and uniform nuclear matter
\cite{ser1,rin1,ser2}. These effective field theories are based on a
Lagrangian density that contains the nucleon (as an elementary Dirac
particle) together with an isoscalar-scalar ($\sigma$) meson, an
isoscalar-vector ($\omega$) meson, an isovector-vector ($\rho$) meson,
and the photon as the relevant degrees of freedom for describing the
nuclear many-body problem. In the mean field approximation the meson
fields are replaced by their ground-state expectation values, thereby
becoming classical fields. The quantum structure of the theory is
carried by the nucleon field. For systems with time reversal symmetry,
only the time-like component of the vector fields contribute. At the
mean field (``tadpole'') level, charge conservation implies that only
the third component (in isospin space) of the isovector $\rho$-meson
field does not vanish. As a final product, one obtains the following
Hamiltonian density:
%
%% Equation 1
\begin{eqnarray}
\label{PV1}
{\cal H}({\bf r}) & = &  \sum_\alpha \varphi_\alpha^\dagger
\left [ -i \mbox{\boldmath$\alpha$} \!\cdot\! \mbox{\boldmath$\nabla$}
+ \beta (M - \Phi) + W
+ \frac{1}{2}\tau_3 B
+ \frac{1+\tau_3}{2} A\right ] \varphi_\alpha
\nonumber \\[3mm]
& &
+\frac{1}{2g_{\rm s}^2} \left [\left (\mbox{\boldmath$\nabla$}\Phi\right )^2
+m_{\rm s}^2\Phi^2\right ] +\frac{\kappa}{3!}{\Phi^3}+\frac{\lambda}{4!}{\Phi^4}
-\frac{1}{2g_{\rm v}^2} \left [\left (\mbox{\boldmath$\nabla$} W\right )^2
+m_{\rm v}^2 W^2\right ]
\nonumber \\[3mm]
& &
-\frac{1}{2g_\rho^2} \left [\left (\mbox{\boldmath$\nabla$} B\right )^2
+m_\rho^2 B ^2\right ]
-\Lambda_{\rm v} W^2B^2
-\frac{1}{2} \left (\mbox{\boldmath$\nabla$} A\right )^2,
\end{eqnarray}
where the summation runs over the occupied nucleon states
$\varphi_\alpha ({\bf r})$ of positive energy and $\tau_3$ stands for
the third component of the isospin operator. The scaled meson fields
associated with the $\sigma$, $\omega$, and $\rho$ mesons are,
respectively, $\Phi\equiv g_{\rm s}\phi_0({\bf r})$, $W\equiv g_{\rm
v}V_0({\bf r})$, and $B\equiv g_\rho b_0({\bf r})$. Finally, $A\equiv
e A_0({\bf r})$ represents the time-like component of the photon
field. 

The variation of Eq.~(\ref{PV1}) with respect to the Dirac spinors
yields the Dirac equation satisfied by the nucleons, while the
variations with respect to the meson fields lead to the Klein-Gordon
equations for the corresponding mesons \cite{ser1,rin1,ser2}. The
functional in Eq.~(\ref{PV1}) contains cubic and quartic scalar meson
self-interactions (couplings $\kappa$ and $\lambda$, respectively)
that are tuned to bring the value of the compression modulus of
symmetric nuclear matter down to the empirical value
($K\!=\!200$--300~MeV)~\cite{bog1}. Further, through the inclusion
of these terms one can accurately reproduce the systematics of finite
nuclei throughout the periodic table as, e.g., in the case of the
celebrated relativistic mean field (RMF) parameter set
NL3~\cite{lal1}. The additional isoscalar-isovector cross coupling
$\Lambda_{\rm v}$ enables one to modify the density dependence of the
symmetry energy without compromising the success of the model.

Insofar as we are interested in the properties of some relatively
heavy nuclei that have been identified as possible candidates for APNC
experiments, we extend the model of Ref.~\cite{tod1} to include
nuclear deformation and pairing correlations in the present RMF study.
Only quadrupole deformation ($\beta_2$) is included though, as it is
known that hexadecapole deformation is generally small for the nuclei
under consideration~\cite{vre1}. A BCS framework is employed to deal
with the pairing correlations. Note that because the nuclei to be
studied lie relatively close to the $\beta$-stability line, the use of
the BCS approximation is justified as the Fermi level is sufficiently
bound and the mixing with continuum states---essential for drip line
nuclei---is not relevant here. As in most BCS calculations, the
strength of the neutron-neutron and proton-proton pairing interactions
is determined from the neutron and proton pairing gaps (respectively,
$\Delta_n$ and $\Delta_p$) evaluated from the experimental odd-even
mass differences, using a five-point formula~\cite{ben0}.

The Dirac equation and the Klein-Gordon equations for the meson fields
are treated using an axially deformed basis expansion method as
described in detail in Refs.~\cite{rin2,sil1}. The calculations
presented here have been carried out by expanding the fermionic wave
functions and the bosonic fields in 12 and 20 oscillator shells,
respectively. The solution of the equations in an axially deformed
basis for the odd-even and odd-odd nuclei is more complicated. Indeed,
in these cases time reversal symmetry is broken and the odd particle
induces polarization currents and time-odd components in the mean
fields. However, the impact of these effects on deformation and
binding energies is small~\cite{lal2} and will be neglected
henceforth. In the pairing calculation of the odd nucleon we use the
blocking approximation~\cite{rin3}, thereby restoring time reversal
invariance in the intrinsic frame.

For the isotopic chains under consideration for APNC experiments,
namely, Cs, Ba, Dy, and Yb, the NL3 parameter set yields ground-state
properties, such as binding energies and deformations, in good
agreement with the available experimental data~\cite{vre1}. For our
calculations we adopt the NL3 parameter set and suitably modify it
with the addition of the isoscalar-isovector coupling $\Lambda_{\rm
v}$, prompted by the fact that the slope of the nuclear symmetry
energy at (or near) the saturation density is at present unknown.
Thus, we utilize the nonlinear isoscalar-isovector term $\Lambda_{\rm
v}$ to change the density dependence of the symmetry energy, which in
turn modifies the neutron radius of neutron-rich nuclei. This can be
done without a significant change in those ground-state properties of
finite nuclei that are well constrained experimentally~\cite{hor1}. In
practice, for a given $\Lambda_{\rm v}$ value we readjust the
nucleon-rho coupling constant $g_\rho$ so that the symmetry energy of
nuclear matter at a density of $\rho\!=\!0.10~{\rm fm}^{-3}$ be 25.68
MeV~\cite{tod1,hor1,hor2}. The choice is motivated by the fact that
the symmetry energy of uniform nuclear matter at saturation density is
not well constrained by the known properties of finite nuclei. Rather,
it is some average between the bulk symmetry energy at saturation and
the surface symmetry energy which is constrained by the binding energy
of finite nuclei~\cite{bro1,furns1}. For example, typical effective
nuclear forces fitted to the empirical energies of nuclei predict a
liquid drop model symmetry energy coefficient, which contains a
surface correction, of 22--24~MeV for $^{208}$Pb~\cite{cen1}, far
smaller than the value of the bulk symmetry energy coefficient for
that interaction.

\begin{figure}
\includegraphics[width=0.85\linewidth, angle=0, clip=true]{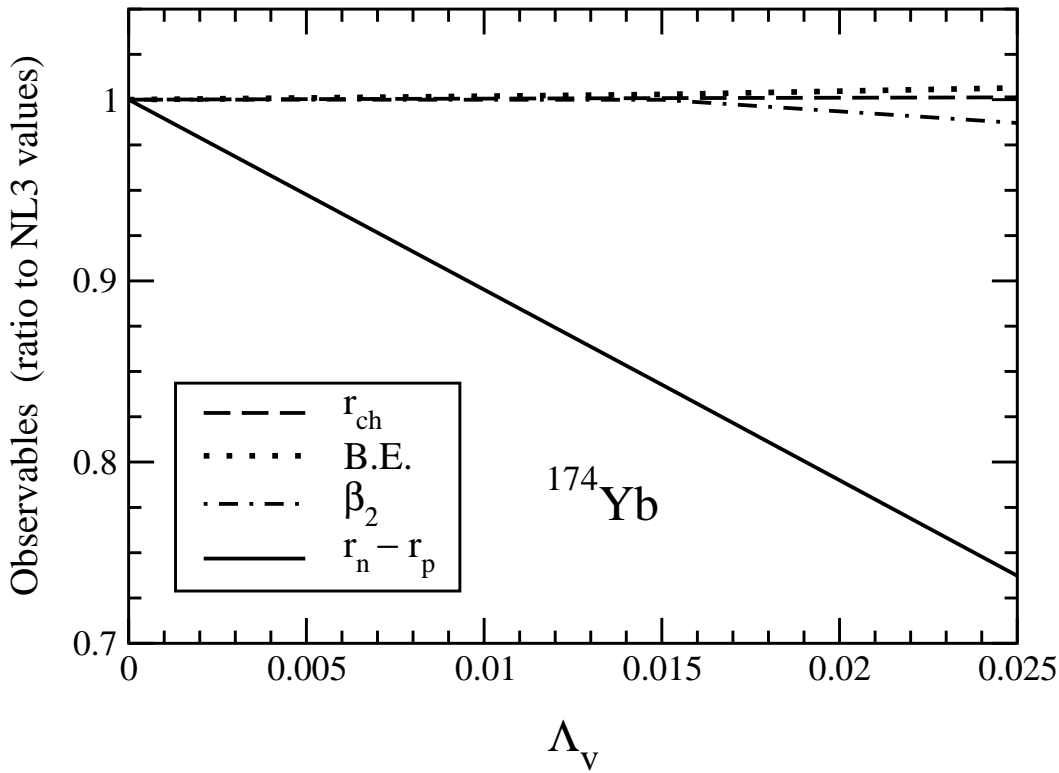}
\caption{\label{Fig1} The dependence of the ground-state charge
radius, binding energy, quadrupole deformation, and neutron skin on
the isoscalar-isovector coupling $\Lambda_{\rm v}$ is exemplified for
the $^{174}$Yb nucleus. The observables have been normalized to the
values of the NL3 interaction ($\Lambda_{\rm v}\!=\!0$).}
\end{figure}

Although it has been established that a variety of ground-state
properties of spherical nuclei are insensitive to the choice of
$\Lambda_{\rm v}$~\cite{hor1,hor2,tod1}, this is not the case for the
neutron skin. The neutron skin $t \!=\! r_n\!-\!r_p$, where $r_n$ and
$r_p$ are the neutron and proton rms radii ($r_{n(p)} \!\equiv\!
\langle r^{2} \rangle_{n(p)}^{1/2}$), is strongly sensitive to the
density dependence of the symmetry energy and hence, to the
isoscalar-isovector coupling $\Lambda_{\rm v}$. This is because the
slope (i.e., pressure) of the symmetry energy controls how
far is the neutron skin pushed out relative to the symmetric core. In
order to show that this scenario holds true even when the spherical
symmetry is broken and pairing correlations are important, we display
in Fig.~\ref{Fig1} the charge radius, binding energy, quadrupole
deformation, and the neutron skin as a function of $\Lambda_{\rm v}$
for the representative heavy nucleus ${}^{174}$Yb. Note that all
observables in Fig.~\ref{Fig1} have been normalized to the
corresponding NL3 values ($\Lambda_{\rm v}\!=\!0$). It is seen that
the charge radius, binding energy and quadrupole deformation are
insensitive to the value of $\Lambda_{\rm v}$ in the 0--0.025 range;
the quadrupole deformation displays the largest change, but this
amounts to only 2\%. In contrast, the neutron skin decreases rapidly
with increasing $\Lambda_{\rm v}$ and it varies by about 
25\% when $\Lambda_{\rm v}$ is changed from 0 to 0.025. This reflects
the softening of the symmetry energy as $\Lambda_{\rm v}$ increases.
Therefore, we conclude that neither pairing correlations nor
deformation alter our earlier conclusions about the role of
$\Lambda_{\rm v}$ in the systematics of finite nuclei.

\begin{figure}
\includegraphics[width=1.0\linewidth, angle=0, clip=true]{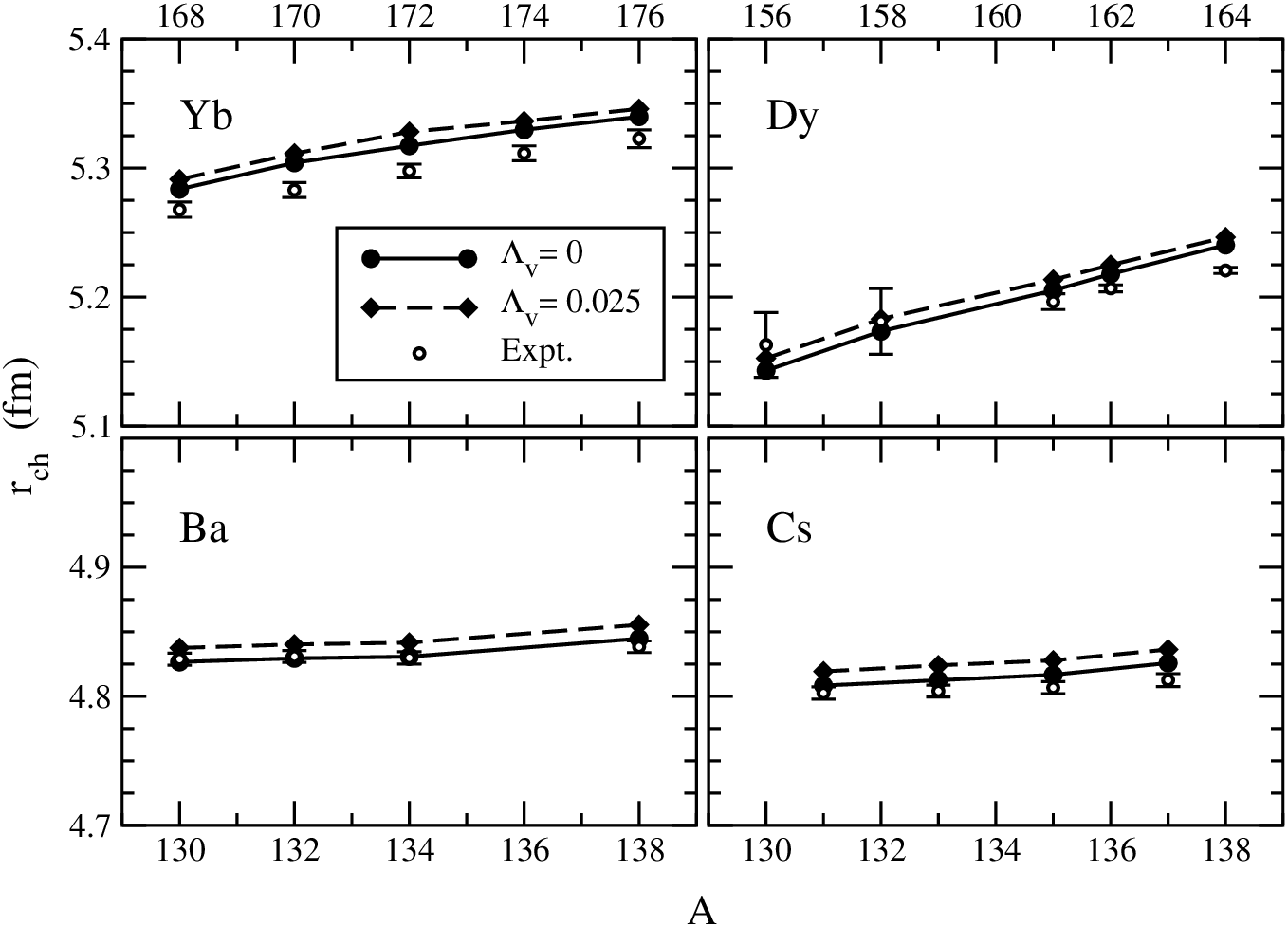}
\caption{\label{Fig2} Charge radii calculated with the models
considered in the text are compared with the available experimental
data~\cite{ang1} in isotope chains of interest for APNC studies. The
actual experimental error bar for $^{156}$Dy and $^{158}$Dy has been
reduced by a factor of 10 for display.}
\end{figure}

The neutron skin of a nucleus is an important component in the study
of atomic parity non-conservation. So is its charge radius. While
neutron radii of heavy nuclei ($r_n$) are poorly known, high precision
data for the charge radii are available. To test the reliability of
our model, we now compare in Fig.~\ref{Fig2} calculated charge radii
with experimental data for some of the relevant isotopes
used in APNC experiments; these include Cs, Ba, Dy and Yb. 
The experimental values are taken from the recent compilation of
Ref.~\cite{ang1}, which combines measured data from electron
scattering, from muonic X-rays, and from $K_\alpha$ and optical
isotope shifts for the purpose of providing a unified set of nuclear
charge radii. Theoretical calculations are presented with the original
NL3 parametrization ($\Lambda_{\rm v}\!=\!0$) and with one for which
the density dependence of the symmetry energy has been softened by
fixing the value of the isoscalar-isovector coupling to $\Lambda_{\rm
v}\!=\!0.025$. It can be seen that both sets of calculated charge
radii are in excellent agreement with the experimental values for {\em
all} of the considered nuclei. At a more quantitative level, we find
that the relative ${\chi}^2$ value between the theoretical and
experimental charge radii for the isotopic chains of Cs, Ba, Dy, and
Yb is always smaller than 0.25\% for both ($\Lambda_{\rm v}\!=\!0$ and
$\Lambda_{\rm v}\!=\!0.025$) parameter sets. This result validates the
use of both parameter sets in computing key nuclear observables of
relevance to the APNC program. Note that the binding energies and
deformations for all the nuclei considered here are also in good
agreement with the experimental data (not shown). Moreover, and as
alluded earlier, these observables are highly insensitive to the
choice of $\Lambda_{\rm v}$.

\begin{figure}
\includegraphics[width=1.0\linewidth, angle=0, clip=true]{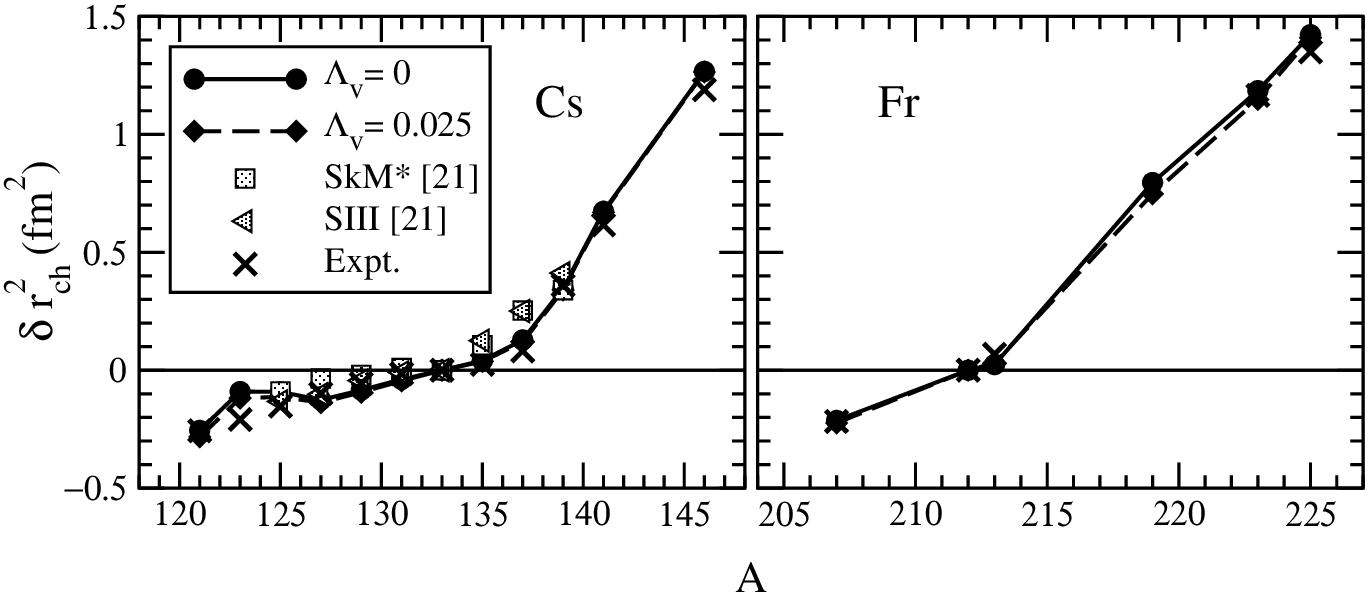}
\caption{\label{Fig2new} Isotope shifts of the charge radii of cesium
and francium nuclei. The experimental data are from Ref.~\cite{ang1}.}
\end{figure}

A more daring challenge for a theoretical model than the charge radii
themselves is the reproduction of the observed charge isotope shifts.
One famous case is the pronounced kink exhibited by the experimental
charge radii of the Pb nuclei at the magic neutron number $N=126$. The
kink is not reproduced by conventional non-relativistic theories
(either Skyrme or Gogny forces). In contrast, the RMF calculations
show a remarkable agreement with experiment \cite{rin1}. The origin
for the deviations between the two models has been traced back,
largely, to the isospin dependence of the spin-orbit term. We present
in Fig.~\ref{Fig2new} results for the isotope shifts in two of our
flagship chains for the subsequent APNC analysis: the cesium and
francium alkali metals. The shifts are referred to $^{133}$Cs and
$^{212}$Fr, respectively, as are the data of Ref.~\cite{ang1}. The
experimental method of preference for the measurement of nuclear
charge isotope shifts is high-resolution laser spectroscopy, which
exploits the isotopic effects on the hyperfine structure of atomic
optical transitions. In addition to the RMF calculations with
$\Lambda_{\rm v}=0$ and $\Lambda_{\rm v}=0.025$ we have drawn in
Fig.~\ref{Fig2new} the SkM* and SIII results of the non-relativistic
Hartree-Fock calculations of Ref.~\cite{che1} for Cs. One immediately
recognizes the ability of the RMF model in reproducing the trend of
the changes of the experimental data when neutrons are
added; conspicuously, the change of slope at the $N=82$ shell closure
for Cs and at the $N=126$ shell closure for Fr. In spite of this, the
theoretical predictions may be off the experiment by a factor of $\sim
2$ in some specific isotopes. One should keep in mind that this
happens in cases where the isotope shift is a small quantity and that
it emerges from the cancellation of two much larger numbers (the
square charge radii of two isotopes). Similarly to the observables
discussed above, the calculated isotope shift is a quantity that
remains unaffected by the $\Lambda_{\rm v}$ coupling.

As has been mentioned in earlier publications~\cite{typ1,cen2,furns1},
typical relativistic parametrizations predict a neutron skin in
$^{208}$Pb of about $0.3$~fm. This should be contrasted with the case
of non-relativistic forces of the Skyrme or Gogny type that yield
values that are significantly smaller (between 0.1 and 0.2~fm). As the
proton rms radius in $^{208}$Pb is reproduced by both relativistic and
non-relativistic models, the spread in the neutron rms radius in
$^{208}$Pb---owing to the different density dependence of the symmetry
energy in different theoretical models---is about 0.1--0.2~fm. Within
the relativistic mean-field models a spread (i.e., a theoretical
uncertainty) of the neutron skin for a specific nucleus can be
simulated in a controlled manner through the addition of the
isoscalar-isovector coupling $\Lambda_{\rm v}$~\cite{hor1,tod1}.
Starting from the original NL3 parameter set and changing the
isoscalar-isovector coupling constant from $\Lambda_{\rm v}\!=\!0$
(pure NL3 interaction) to $\Lambda_{\rm v}\!=\!0.025$, the neutron skin
in $^{208}$Pb varies from 0.280~fm to 0.209~fm \cite{hor1}. Hence, in
this formalism the induced theoretical uncertainty of the neutron
radius of $^{208}$Pb between these two extreme values of $\Lambda_{\rm
v}$ is approximately $0.07$~fm. This spread for $^{208}$Pb is only
slightly larger than the expected experimental error in the measurement 
of the neutron radius in $^{208}$Pb via the parity violating 
electron-scattering experiment at the Jefferson Laboratory (known as
the PREX experiment)~\cite{mic1}. Thus we take the theoretical spread 
in the neutron radius of $^{208}$Pb as a baseline to estimate a realistic 
uncertainty in those nuclei involved in APNC experiments.

\begin{figure}
\includegraphics[width=0.85\linewidth, angle=0, clip=true]{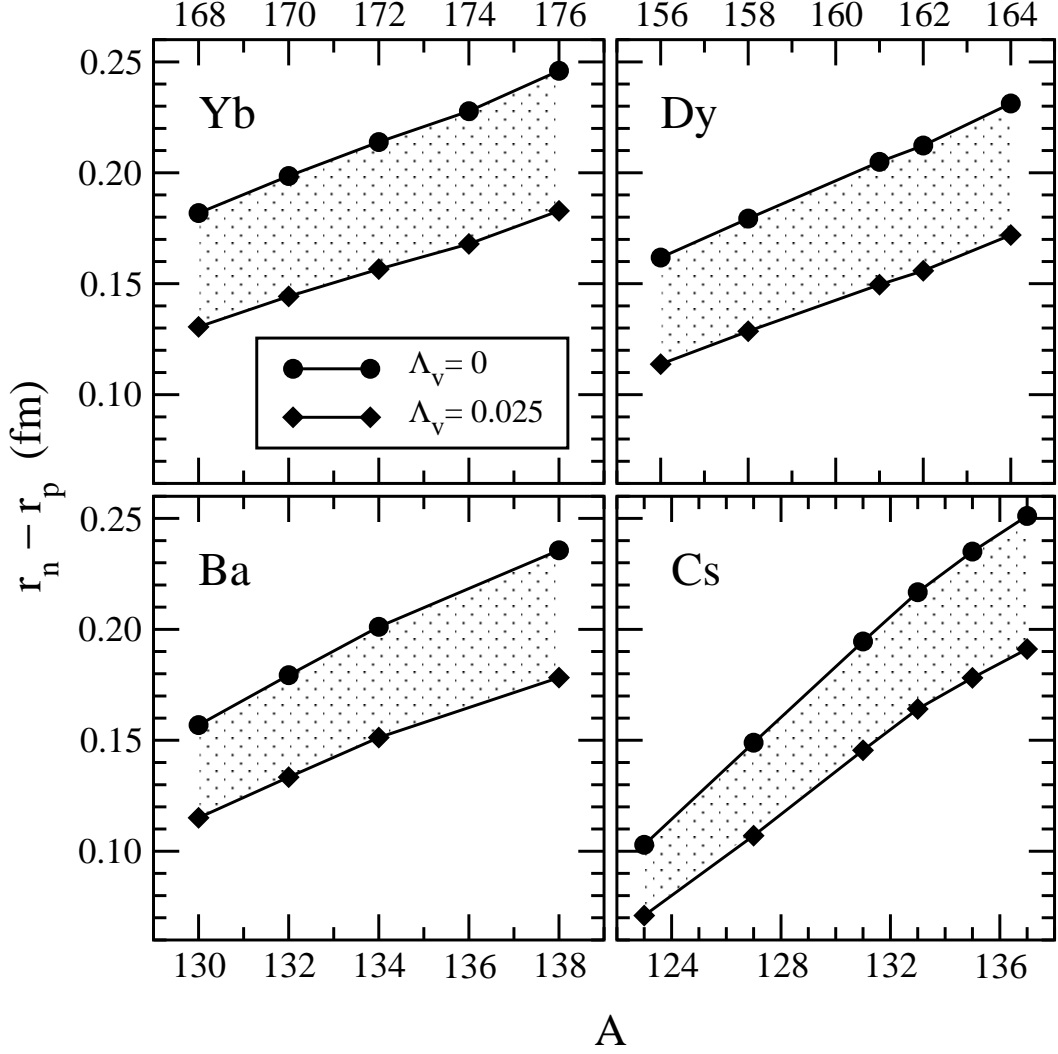}
\caption{\label{Fig3} Variation of the neutron skin in four isotope
chains of possible relevance to APNC isotope ratios measurements, for
the two effective field theory models considered in this work. Note
that the change in the neutron skin of $^{208}$Pb predicted by these
two models, otherwise accurately calibrated for binding energies and
charge radii, mocks up the uncertainty that will be left after the
projected precision measurement of the neutron radius of $^{208}$Pb at
the Jefferson Laboratory~\cite{mic1}.}
\end{figure}

The neutron skins calculated with both of the relativistic
parametrizations (namely, $\Lambda_{\rm v}=0$ and 
$\Lambda_{\rm v}=0.025$) for stable and long-lived isotopes
of Cs, Ba, Dy, and Yb are displayed in Fig.~\ref{Fig3}. For a
specific isotope, the shift between the two curves displayed 
in the figure gives the model spread (due to 
$\Lambda_{\rm v}$) of the neutron skin of the given nucleus. 
For all the considered isotopic chains the neutron skins calculated
with $\Lambda_{\rm v}\!=\!0$ (the stiffest symmetry energy) and
$\Lambda_{\rm v}\!=\!0.025$ (the softest symmetry energy) lie roughly
along two parallel lines. In general, the effect of $\Lambda_{\rm v}$ 
on the neutron skin is more prominent for the more asymmetric and 
heavier nuclei. This can be observed for the relatively long chain of 
Cs isotopes in Fig.~\ref{Fig3}, where the spread of the neutron skin 
varies approximately from 0.03 fm to 0.06 fm in passing from 
$^{123}$Cs to $^{137}$Cs.

\begin{figure}
\includegraphics[width=0.85\linewidth, angle=0, clip=true]{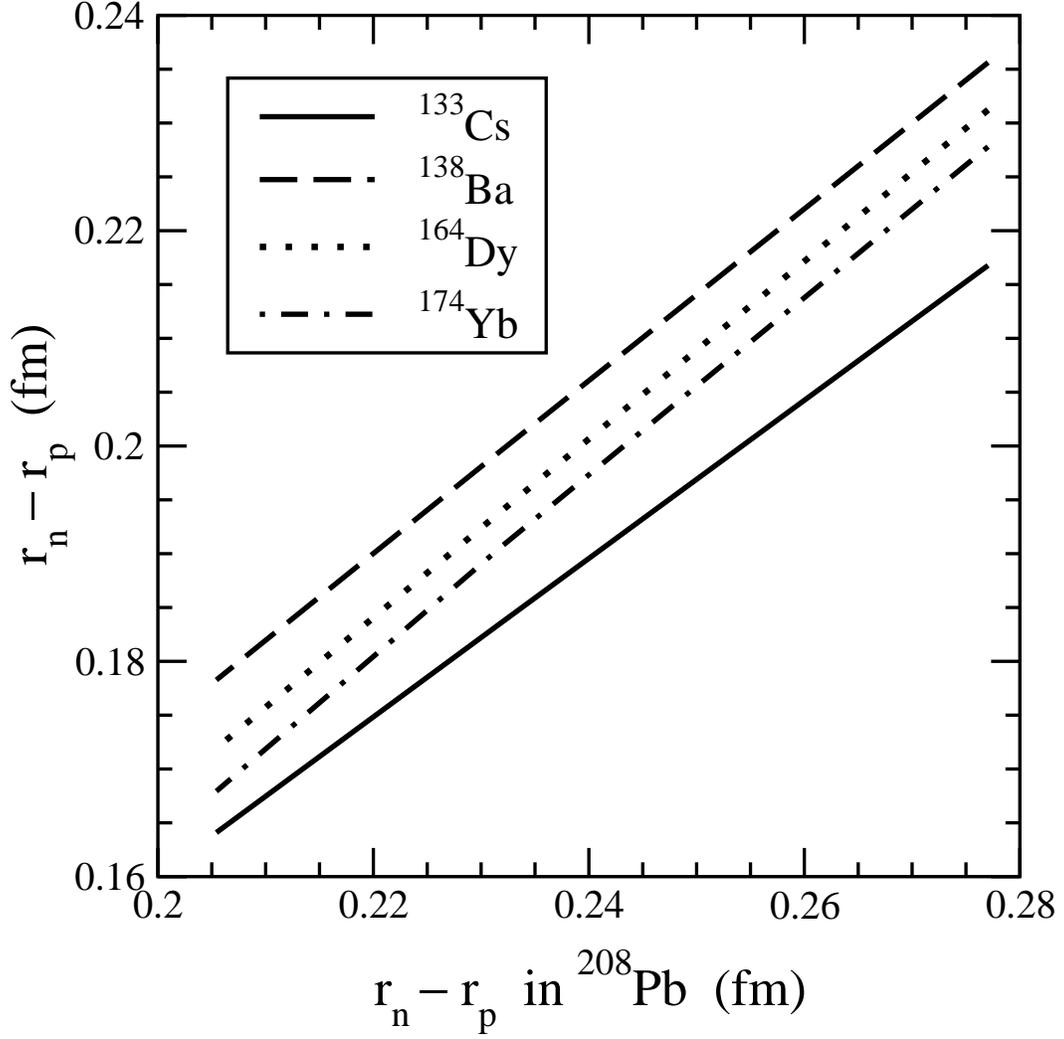}
\caption{\label{Fig4} Model correlation of the neutron skin of
$^{133}$Cs, $^{138}$Ba, $^{164}$Dy, and $^{174}$Yb to the neutron skin
of $^{208}$Pb. The isoscalar-isovector coupling varies between
$\Lambda_{\rm v}=0$ (right side of the figure) and $\Lambda_{\rm
v}=0.025$ (left side).}
\end{figure}

To the extent that in the RMF model used in this work the proton rms
radius of a given nucleus is almost independent of $\Lambda_{\rm v}$,
the theoretical spread of its neutron skin is therefore approximately
equal to the one for its neutron rms radius. At present, the situation
regarding our ignorance of the neutron skin of heavy nuclei is
unlikely to improve because the existing experimental data cannot
reduce the spread in the neutron radii of Cs, Ba, Dy, or Yb any
further. Thus, we now examine whether the PREX experiment on the
neutron radius of $^{208}$Pb may place important constraints on the
neutron skin of APNC nuclei, including those deformed isotopes with an
odd number of nucleons. Figure~\ref{Fig4} evidences the strong linear
correlation~\cite{bro1,typ1,tod1} between the neutron skin of
$^{133}$Cs, $^{138}$Ba, $^{164}$Dy, and $^{174}$Yb and that of
$^{208}$Pb. The calculations include values for the
isoscalar-isovector coupling constant $\Lambda_{\rm v}$ ranging from 0
to 0.025. Note that the correlation holds even for the case of
deformed nuclei having an odd number of nucleons. With the culmination
of the PREX experiment, the present theoretical spread in the neutron
radius of $^{208}$Pb of $\sim 0.3$~fm (relativistic and
non-relativistic models altogether) \cite{furns1,pol2,tod1} will be
replaced by a genuine experimental error five times smaller; that is,
$\Delta r_n(^{208}{\rm Pb})\approx 0.056$~fm~\cite{tod1}. This
experimental error in the neutron radius of $^{208}$Pb will also be
the experimental error in the neutron skin, as the proton radius is
known with a much higher accuracy. If the expected experimental
accuracy is attained, then the spread in the neutron skin of the
several APNC isotopes will also be (indirectly) reduced. In turn, this
result will impact favorably on APNC observables by appreciably
reducing the nuclear structure uncertainty.

% >>>>>>>>>>>>>>>>>>>>>>>>>>>>>>>>>>>>>>>>>>>>>>>>>>>>>>>>>>>>>>>>>>>>

\section{Results for APNC observables}
\label{Results}

Parity violation in atomic systems arises from the interference
between the parity conserving electromagnetic interaction and the
parity violating weak interaction. Although most of the binding energy
of the atomic electrons comes from their attractive Coulomb
interaction with the $Z$ protons in the nucleus, the weak charge of
the nucleus $Q_W$ induces a small correction in the binding energy and
parity of the electronic wave functions. These small corrections can
be measured experimentally. The experimentally measured quantity is
the electric dipole amplitude between two electronic states that in
the absence of parity violating effects would have the same parity,
thereby making the amplitude vanish. The observable related to the
dipole amplitude can be parametrized as 
follows~\cite{pol2,che1,ram1,pan1,pol1}:
\begin{equation}
\label{PV2}
 A_{\rm PNC}(N,Z) = \xi(Z) Q_{W}(N,Z) \equiv
                    \xi(Z) \Big[Q_W^{\rm SM}(N,Z)+
                           \Delta Q^{n-p}_{W}(N,Z)\Big]\;,
\end{equation}
where $\xi(Z)$ embodies the atomic structure contribution and 
$Q_{W}(N,Z)$ represents the weak charge of a nucleus of neutron 
number $N$ and electric charge $Z$. 

The experimentally measured weak charge of the nucleus (without
radiative corrections~\cite{pol2}) $Q_{W}(N,Z)$ differs from the
standard model prediction,
\begin{equation}
\label{PV4}
 Q_W^{\rm SM}(N,Z)=-N+Z(1-4\sin^{2}\theta_{W})\;,
\end{equation}
by a nuclear correction factor that arises from the difference 
between proton and neutron one-body densities. That is,
\begin{equation}
\label{PV4b}
\Delta Q^{n-p}_W(N,Z) = N \left(1 - \frac{q_n}{q_p}\right)\;,
\end{equation} 
with
\begin{equation}
\label{PV6}
 q_{n(p)}=\int d {\bf r} f(r)\rho_{n(p)}(r) \;.
\end{equation}
Note that in Eq.~(\ref{PV4}) $\theta_{\rm W}$ is the Weinberg (or weak
mixing) angle, while $\rho_{n(p)}$ in Eq.~(\ref{PV6}) is the
neutron (proton) density normalized to one, and $f(r)$ is an
electronic form factor that describes the spatial variation of the
electronic axial-vector matrix element over the size of the nucleus. 

The atomic parity formalism ``starts'' by assuming that neutrons and
protons have the same spatial distribution and then adds a
nuclear-structure correction factor to quantify the differences
between the actual neutron and proton densities. This is motivated by
the fact that the proton density, which is needed to determine various
parity conserving observables, is very accurately known.  During the
course of this presentation---and as has been done
elsewhere~\cite{pol2,che1}---an additional quantity that parametrizes
the nuclear structure corrections will be extensively used. This
quantity is defined as follows:
\begin{equation}
\label{PV5}
 Q_W^{\rm nucl}(N,Z)=-N(q_n-1)+Z(1-4\sin^{2}\theta_{W})(q_p-1)\;.
\end{equation}
Note that, in terms of this quantity, the weak charge of the nucleus
may be written as
\begin{eqnarray}
\label{PV7}
   Q_{W}(N,Z) &=& \Big[Q_W^{\rm SM}(N,Z)+ \Delta Q^{n-p}_{W}(N,Z)\Big]
                  \nonumber \\ 
              &=& \Big[-Nq_n+Zq_{p}(1-4\sin^{2}\theta_{W})\Big] 
                  \frac{1}{q_{p}}
                  \nonumber \\ 
              &=& \Big[Q_W^{\rm SM}(N,Z)+Q_W^{\rm nucl}(N,Z)\Big]
                  \frac{1}{q_{p}}
                  \;.
\end{eqnarray}

To compute the axial form factor $f(r)$ one expands the electronic
Dirac wave functions in a power series about the origin~\cite{pol2}
(the Coulomb potential is assumed to be that of a uniform nuclear
charge distribution). Using this prescription yields the electronic
form factor in closed form:
\begin{equation}
\label{PV9}
f(r)=1-\frac{(Z\alpha)^2}{2}\left[\left(\frac{r}{R_p}\right)^2
-\frac{1}{5}\left(\frac{r}{R_p}\right)^4
+\frac{1}{75}\left(\frac{r}{R_p}\right)^6\right],
\end{equation}
where $R_p$ is the cutoff radius of a sharp nuclear charge density.
Effects from the finite nuclear size are computed from the
ground-state expectation value of the time-like component of the weak
vector current by considering the neutron and proton as point-like
particles.
Again, we assume sharp proton and neutron densities with rms radii
equal to those predicted by self-consistent RMF models. Note that the
relation between the sharp cutoff radii $R_{n(p)}$ and the
corresponding rms radii $r_{n(p)}$ is given by
\begin{equation}
 R_{n(p)}=\sqrt{\frac{5}{3}} \; r_{n(p)}\;.
 \label{Rpn}
\end{equation}
Estimates for the weak matrix elements $q_p$ and $q_n$ are then given
by the following expressions:
\begin{eqnarray}
\label{PV10}
q_p&=& 1- \frac{817}{3150}(Z\alpha)^2,\\
\label{PV11}
q_n&=&1-(Z\alpha)^2\left[\frac{3}{10}\left(\frac{R_n}{R_p}\right)^2
-\frac{3}{70}\left(\frac{R_n}{R_p}\right)^4
+\frac{1}{450}\left(\frac{R_n}{R_p}\right)^6\right]
\nonumber\\
&=&1-(Z\alpha)^2\left[\frac{817}{3150}+\frac{232}{525}
\frac{t}{r_p}+{\cal O}\Big((t/r_p)^{2}\Big)\right].
\end{eqnarray}
Although in the present approximation $q_p$ does not depend
on nuclear structure effects, the sensitivity of $q_n$ to nuclear
structure uncertainties appears in the form of the rms proton
radius---which is well known---and the poorly known neutron skin $t$.

% >>>>>>>>>>>>>>>>>>>>>>>>>>>>>>>>>>>>>>>>>>>>>>>>>>>>>>>>>>>>>>>>>>>>
%RATIO ANALYSIS

\subsection{Prospects for isotope ratios analyses}
The measurement of the weak charge in APNC experiments is plagued by
theoretical uncertainties in both atomic and nuclear structure.
Fortunately, uncertainties in atomic structure may be eliminated, or
at least considerably reduced, by studying parity violation along a
chain of isotopes and taking ratios of APNC
measurements~\cite{Dzuba86,for1,pol2,che1,ram1}. According to
Eq.~(\ref{PV2}), the dependence of the parity violating amplitude
$A_{\rm PNC}$ on the atomic theory contribution $\xi(Z)$ will cancel
out in the ratio of two measurements performed in two different
isotopes of the same element, provided $\xi(Z)$ does not change
appreciably along that isotopic chain. Thus, we now address the role
of the remaining---nuclear structure---uncertainty in APNC observables
along the isotopic chains of Cs, Ba, Dy, and Yb. The observables of
interest are expressed in the form of the following two ratios:
\begin{eqnarray}
\label{PV3a}
{\cal R}_1&=&\frac{Q_W(N^\prime,Z)-Q_W(N,Z)}
                  {Q_W(N^\prime,Z)+Q_W(N,Z)}\nonumber\\
&=&\frac{Q_W^{\rm SM}(N^\prime,Z)+\Delta Q^{n-p}_W(N^\prime,Z)-
         Q_W^{\rm SM}(N,Z)- \Delta Q^{n-p}_W(N,Z)}
        {Q_W^{\rm SM}(N^\prime,Z) + \Delta Q^{n-p}_W(N^\prime,Z)+
         Q_W^{\rm SM}(N,Z)+ \Delta Q^{n-p}_W(N,Z)}\nonumber\\
&\approx&
   \frac{Q_W^{\rm SM}(N^\prime,Z)+ Q_W^{\rm nucl}(N^\prime,Z)-
         Q_W^{\rm SM}(N,Z)- Q_W^{\rm nucl}(N,Z)}
        {Q_W^{\rm SM}(N^\prime,Z) + Q_W^{\rm nucl}(N^\prime,Z)+
         Q_W^{\rm SM}(N,Z)+ Q_W^{\rm nucl}(N,Z)} \;,
\end{eqnarray}
and
\begin{eqnarray}
\label{PV3}
 {\cal R}_2=\frac{Q_W(N^\prime,Z)}{Q_W(N,Z)}&=&
            \frac{Q_W^{\rm SM}(N^\prime,Z)+ \Delta Q^{n-p}_W(N^\prime,Z)}
                 {Q_W^{\rm SM}(N,Z)+\Delta Q^{n-p}_W(N,Z)}\nonumber\\
            &\approx&
            \frac{Q_W^{\rm SM}(N^\prime,Z)+ Q_W^{\rm nucl}(N^\prime,Z)}
                 {Q_W^{\rm SM}(N,Z)+ Q_W^{\rm nucl}(N,Z)}\;.
\end{eqnarray}
The approximate sign ($\approx$) in the above two equations follows
from the assumption that the overlap $q_{p}$ appearing in 
Eq.~(\ref{PV7}) remains constant along the whole isotopic chain.
As can be realized from Tables~\ref{Table2} and ~\ref{Table3} (see
below), this is an excellent approximation.

It has been argued in Ref.~\cite{ram1} that corrections to standard
model predictions induced by new physics or uncertainties in nuclear
structure are essentially the same whether one uses ${\cal R}_1$ and
${\cal R}_2$. Thus, we focus here exclusively on ${\cal R}_1$. Using
Eqs.~(\ref{PV4}) and (\ref{PV5}) we can write the ratio ${\cal R}_1$
approximately as
\begin{equation}
\label{PV8}
{\cal R}_1 \approx \frac{Q_W^{\rm SM}(N^\prime,Z)-Q_W^{\rm SM}(N,Z)}{Q_W^{\rm SM}
(N^\prime,Z)
   +Q_W^{\rm SM}(N,Z)}\left[1+\frac{N^\prime}{\Delta N}
   [ q_n(N^\prime,Z) - q_n(N,Z) ] \right],
\end{equation} 
where $N^{\prime}$ ($N$) is the largest (smallest) neutron number and
$\Delta N \!\equiv\! N' \!-\! N$ represents the difference in neutron
number between the two extremes of the isotopic chain. Further,
Ref.~\cite{ram1} also established that a significant test of the
standard model on the basis of isotope ratios requires a measurement
of ${\cal R}_1$ with an accuracy better than $0.1\%$. Otherwise, a
less precise determination of isotope ratios would not be worthwhile
to compete with the sensitivity to new physics of the presently known
data for the cesium isotope $^{133}$Cs. The measurement of the
electron-nucleon parity violating effect in $^{133}$Cs to 0.35\%
accuracy \cite{Wood97} remains to date as the paradigm of APNC
experimental precision.

Because atomic uncertainties have been eliminated from the ratio
(\ref{PV8}), the remaining uncertainties in ${\cal R}_1$ reflect the
known accuracy in the neutron and proton rms radii. While proton
densities have been determined from electron-scattering experiments
with remarkable accuracy, knowledge of the neutron densities to a
comparable level of accuracy is lacking. Thus, the main nuclear
structure uncertainty in the isotopic ratio ${\cal R}_1$ comes from
our poor knowledge of the neutron radii (or neutron skin) of heavy
nuclei. Indeed, by inserting Eq.~(\ref{PV11}) into Eq.~(\ref{PV8}),
the relative uncertainty in ${\cal R}_1$ may be approximated by
\begin{equation}
\label{PV13}
\frac{\delta {\cal R}_1}{{\cal R}_1} \approx 
-\frac{232}{525}(Z\alpha)^2\frac{N^\prime}{\Delta N} \, \delta\!
\left[ \frac{t}{r_p}(N^\prime,Z)
- \frac{t}{r_p}(N,Z)\right]\;.
\end{equation}

In Sec.~\ref{Theory} we showed that a spread in the neutron radius
of $^{208}$Pb is obtained by using parametrizations with
isoscalar-isovector couplings $\Lambda_{\rm v}\!=\!0$ and
$\Lambda_{\rm v}\!=\!0.025$, which is representative of the
experimental error ($\sim 1\%$) expected from the Jefferson Laboratory
experiment. Note that the quoted error in the neutron radius of
$^{208}$Pb is also the error in its neutron skin as the proton radius
is known with a much higher accuracy. Inasmuch as a linear correlation
between the neutron skin of $^{208}$Pb and that of the APNC nuclei has
been established (see Fig.~\ref{Fig4}), an estimate of the theoretical
uncertainty in the weak charge can be obtained from the spread in the
neutron skin of the APNC nuclei using the $\Lambda_{\rm v}\!=\!0$ and
the $\Lambda_{\rm v}\!=\!0.025$ parameter sets.

Let us outline the procedure we shall follow to determine the relative
error in ${\cal R}_1$ (namely, $\delta {\cal R}_1/{\cal R}_1$). First,
we calculate the neutron skin $t\!=\!r_{n}\!-\!r_{p}$ in a certain
isotopic chain using the $\Lambda_{\rm v}\!=\!0$ parametrization.
Since $r_p$ is roughly similar for all the isotopes of the chain, the
fractional neutron skin variation along the chain is estimated as
\begin{equation}
 (\Delta\bar{t})_{\Lambda_{\rm v}=0}= 
 \frac{t(N',Z)-t(N,Z)}{\langle{r_p}\rangle}\Big|_{\Lambda_{\rm v}=0}\;, 
\end{equation}
where $(N',Z)$ refers to the heaviest member and $(N,Z)$ to the
lightest member of the isotopic chain under consideration, and
$\langle{r_p}\rangle$ is the average proton rms radius of the nuclei
in the chain. Next, we compute the corresponding fractional neutron
skin variation $(\Delta\bar{t})_{\Lambda_{\rm v}=0.025}$ for the
softer $\Lambda_{\rm v}\!=\!0.025$ parameter set. Finally, the model
spread \cite{ram1} is obtained as the difference between the
fractional neutron skin variations calculated in the two parameter
sets. That is,
\begin{equation}
\delta (\Delta\overline{t})=(\Delta\bar{t})_{\Lambda_{\rm v}=0}
                           -(\Delta\bar{t})_{\Lambda_{\rm v}=0.025}\;.
 \label{ModelSpread}
\end{equation}

In Table~\ref{Table1} we display the relative uncertainty $\delta
{\cal R}_1/{\cal R}_1$ in the ratio ${\cal R}_1$ for each one of the
isotope chains (Cs, Ba, Dy, and Yb) together with the different
components required to compute it. These are the largest neutron
number of the isotopic chain $N'$, the largest difference in the
neutron number of the isotopic chain $\Delta N$, the average proton
rms radius $\langle{r_p}\rangle$, and the model spread
$\delta(\Delta\bar{t})$. The latter turns out to be a small quantity,
not only in the present relativistic calculations but also in
calculations with non-relativistic nuclear energy functionals (see
Ref.~\cite{ram1} and references quoted therein). The reason resides in
the approximately linear relationship of the neutron skin along a
chain of isotopes with the neutron number. While the model dependence
in the value of the intercept is obvious (see Fig.~\ref{Fig3}), the
slope is only mildly model dependent. And it is the difference in
slope that the model spread $\delta(\Delta\bar{t})$ is sensitive to.

It is seen that the estimated relative uncertainty of the APNC ratios
${\cal R}_1$ in the four isotopic chains considered is around
0.25\%--0.35\%, with a slight increase with increasing atomic
number. The quoted uncertainty appears somewhat larger than the
$0.1\%$ value that would be desirable to extract new physics limits
beyond the present reach of the $^{133}$Cs experiment. Note that the
neutron skin along a given isotopic chain varies (almost) linearly
with neutron number (see Fig.~\ref{Fig3}). This makes the model spread
$\delta(\Delta\bar{t})$ roughly proportional to the neutron difference
$\Delta N$. As the combination $\delta(\Delta\bar{t})/\Delta N$ is
therefore nearly independent of $\Delta N$~\cite{pol1}, it is unlikely
that the precision in ${\cal R}_1$ can be improved by enlarging the
range of the neutron difference $\Delta N$ between the two extremes of
the isotopic chain [see Eq.~(\ref{PV13})].

Table~\ref{Table1} also displays the model spread 
$[\delta (\Delta\bar{t})]_{0.1\%}$ that would be required to reach the
sought-after relative uncertainty of 0.1\% in the isotopic ratio
${\cal R}_1$. In all cases this number is smaller than the value
provided by the RMF model.  Recall that we have tuned the effective
interaction by means of $\Lambda_{\rm v}$ to mimic the purported 1\%
accuracy of the PREX experiment.  Hence, it appears that under the
present---and future---situation, nuclear structure uncertainties
affecting the variation of the neutron distribution along an isotopic
chain are too large to make ${\cal R}_1$ a useful probe for physics 
beyond the standard model. However, note that the differences are not 
dramatic: a factor 2.5 for Cs and Ba, 2.75 for Dy, and 3 for Yb. Thus, 
second generation experiments and/or novel facilities may significantly 
aid in this quest.
% >>>>>>>>>>>>>>>>>>>>>>>>>>>>>>>>>>>>>>>>>>>>>>>>>>>>>>>>>>>>>>>>>>>>
% SINGLE ISOTOPE ANALYSIS

\subsection{Prospects for single-isotope analyses}

For most of the APNC nuclei---particularly Cs and Fr, with one and no
stable isotopes, respectively---one cannot envisage obtaining
precision data on several members of the isotopic chain from
experiments at the present time. For these cases, and to provide
useful hints for future experiments, accurate estimates of the
uncertainty in the neutron skin are essential. Thus, we now compute
the weak charge ---including nuclear and one-loop electroweak
radiative corrections---for some Cs isotopes using both relativistic
($\Lambda_{\rm v}\!=\!0$ and $\Lambda_{\rm v}\!=\!0.025$)
parametrizations. The relevant formulas to be employed are given by
Eqs.~(\ref{PV4})--(\ref{PV5}) suitably modified by the inclusion of
radiative corrections derived by Marciano and Rosner~\cite{mar1}; that
is, 
\begin{eqnarray}
 &&\hspace{-0.4in}
 Q_W^{\rm SM}(N,Z)=(0.9857 \pm 0.0004)
   \left[-N+Z\Big(1-(4.012\pm 0.010)\bar{x}\Big)
   \right] \;, \label{QWSMRad} \\
 &&\hspace{-0.4in}
 Q_W^{\rm nucl}(N,Z)=(0.9857 \pm 0.0004) 
   \left[-N(q_{n}-1)+Z\Big(1-(4.012\pm 0.010)\bar{x}\Big)
           (q_{p}-1)\right] , \label{QWNucRad}
\end{eqnarray}
where we have adopted the shorthand notation
\begin{equation}
 \bar{x}\equiv\sin^{2}\theta_{W}=0.2323 \;.
 \label{xbar}
\end{equation}
Notice that corrections due to new physics have not been included in
the above expressions; these may be found in Refs.~\cite{mar1,pol2,ram1}. 

To be able to compare more precisely our results with those in the
available literature, the weak matrix elements $q_p$ and $q_n$
appearing in Eq.~(\ref{QWNucRad}) have been slightly modified from
those in Eqs.~(\ref{PV10}) and (\ref{PV11}). We now take into account
the corrections due to the finite size of the proton and neutron
charge distributions, which originate additive correction terms
$q_p^c$ and $q_n^c$ to Eqs.~(\ref{PV10}) and (\ref{PV11}),
respectively. These terms are given by~\cite{pol2}
\begin{equation}
q_{n(p)}^c=\int d {\bf r} \, \frac{1}{6} \, 
  \langle r^2\rangle_{I,n(p)}^{W}
  \, f(r) \, \nabla^2\!\rho_{n(p)}^c / Q_{n(p)}^{W},
\end{equation}
where $\langle r^2\rangle_{I,n(p)}^{W}$ are the intrinsic nucleon
weak rms radii, $\rho_{n(p)}^c$ are the density distributions of
nucleon centers, and $Q_{n(p)}^{W}$ are the nucleon weak charges.
Assuming uniform nucleon distributions, we can cast the corrective
terms $q_{n(p)}^c$ due to the finite size of the nucleon as~\cite{pol2} 
\begin{eqnarray}
\label{QPMod}
q_p^c&=& -(Z\alpha)^2 \frac{0.32}{R^2}
\left(2.1-\frac{0.14}{2Q^{W}_p}\right),\\
\label{QNMod}
q_n^c&=& -(Z\alpha)^2 \frac{0.32}{R^2}
\left(0.74-\frac{0.14}{2Q^{W}_n}\right),
\end{eqnarray}
where $R$ denotes the sharp cutoff charge radius (which now includes
the finite size of the proton) and a contribution of 0.14 fm$^2$ from
the strangeness radius of the nucleon is included.

The analysis of the data of a clean measurement by Wood {\em et al.}
\cite{Wood97} of the amplitude of the parity non-conserving transition
between the 6$s$ and 7$s$ states of $^{133}$Cs, the only naturally
occurring cesium isotope, established a value of $Q^{\rm exp}_W=
-72.06(28)_{\rm exp}(34)_{\rm theor}$ for the weak charge of this
element~\cite{ben1}. This experimentally extracted result differed by
2.5$\,\sigma$ from the standard model prediction with radiative
corrections of $Q^{\rm SM}_W=-73.20 \pm 0.13$ [Ref.~\cite{mar1} and
Eq.~(\ref{QWSMRad})]. Let us mention that $\sigma$ is evaluated by
adding in quadrature the experimental and theoretical errors quoted
for $Q^{\rm exp}_W$. The excitement over this apparent disagreement
has faded as a consequence of recent reports that demonstrate that the
deviation between the experimental results and the standard model
prediction can be removed by taking into account self-energy and
vertex QED radiative corrections~\cite{kuc1,mil1} (see also
Ref.~\cite{Ginges04} for an up-to-date review of the status of
measurements and calculations of parity violation in atoms). The newly
reported values of the weak charge for ${}^{133}$Cs---which reconcile
the experiment with the standard model prediction---are $Q^{\rm
exp}_W=-72.71(29)_{\rm exp}(39)_{\rm theor}$~\cite{kuc1} and $Q^{\rm
exp}_W=-72.90(28)_{\rm exp}(35)_{\rm theor}$~\cite{mil1}. It should be
noted that in the extraction of the experimental values of the weak
charge $Q^{\rm exp}_W$, one assumes equal proton and neutron densities
(normalized to unity), so that the overall nuclear correction (namely
$q_{p}$) factors out. An efficient way to account for the
difference between neutron and proton densities---and thus between
$q_n$ and $q_p$---has been indicated in Eq.~(\ref{PV4b}). Including
radiative corrections, this nuclear structure correction factor
becomes 
\begin{equation}
 \label{PV15}
 \Delta Q^{n-p}_W = 0.9857 N \left(1 - \frac{q_n}{q_p}\right)\;,
\end{equation}
and the total weak charge of the nucleus is computed as $Q_{W}=
Q_W^{\rm SM} + \Delta Q^{n-p}_{W}$.

Results for the weak charge together with other relevant quantities,
such as $q_{n(p)}$ [including the intrinsic structure corrections
(\ref{QPMod}) and (\ref{QNMod})], the proton rms radius $r_p$, and the
ratio $r_n/r_p$ of neutron to proton rms radii are displayed in
Table~\ref{Table2} for Cs isotopes. As expected, we find that $q_p$ is
constant, model independent, and within 0.04\% of the sharp-radius
value expressed in Eq.~(\ref{PV10}). However, both the changes and the
model dependence of the neutron radii along the isotopic chain are
noteworthy. Clearly, the ratio $r_n/r_p$ increases with increasing
neutron number $N$. Further, for a given neutron number $N$, $r_n/r_p$
decreases with increasing $\Lambda_{\rm v}$, owing to the softening of
the symmetry energy. Hence, $q_n$ decreases with increasing neutron
number and with a decreasing $\Lambda_{\rm v}$. This ultimately leads
to an increase in the nuclear part of the weak charge $Q^{\rm nucl}_W$
[see Eq.~(\ref{QWNucRad})]. In summary, while $q_p$ is largely
independent of $\Lambda_{\rm v}$, $q_n$ shows a moderate increase with
$\Lambda_{\rm v}$. Thus, through variations in $\Lambda_{\rm v}$ one
has a margin to adjust the nuclear contribution to the weak charge. In
particular, with the proposed model dependence a change of $0.062$ is
induced in $Q_W^{\rm nucl}$ for the ${}^{133}$Cs atom.

The values of the nuclear structure correction factor $\Delta Q^{n-p}_W$
for the studied Cs isotopes, calculated from (\ref{PV15}), have been
tabulated in Table~\ref{Table2}. For the specific case of ${}^{133}$Cs,
we find a nuclear structure correction factor of $\Delta
Q^{n-p}_W\!=\!0.342$ (0.278) for the $\Lambda_{\rm v}\!=\!0$
($\Lambda_{\rm v}\!=\!0.025$) parameter set. Addition of this nuclear
correction term to the standard model prediction with radiative
corrections ($-73.197$, third column of Table~\ref{Table2}) ends in a
total weak charge
\begin{equation}
 \label{PV15b}
  Q_W(^{133}\text{Cs}) =
 \begin{cases}
   {-72.855}, & \text{if $\Lambda_{\rm v}\!=\!0$,}    \\
   {-72.919}, & \text{if $\Lambda_{\rm v}\!=\!0.025$,}
 \end{cases}
\end{equation}
which is in satisfactory agreement with the measured values according to
the recent revisions~\cite{Ginges04,kuc1,mil1}.

It has been discussed in the literature~\cite{vre1,pan1} that 
$\Delta Q^{n-p}_W$ calculated with RMF parametrizations is almost 
twice as large as the corresponding values predicted by Skyrme 
interactions~\cite{che1}. This is a reflection of the stiffer
symmetry energy of the relativistic models. For example, by using 
the sharp-cutoff-radius approximation of Eqs.~(\ref{PV10}) and
(\ref{PV11}) one can write
%%%
\begin{equation}
\label{PV16}
 \Delta Q^{n-p}_W = 0.9857 N \, \frac{(Z \alpha)^2}{q_p}
                    \, \frac{232}{525} \, \frac{t}{r_p} \;.
\end{equation} 
%%%
Because the above nuclear correction factor is directly proportional
to the neutron skin $t$, RMF parametrizations with stiffer symmetry
energies---and consequently larger neutron skins---yield larger values
for $\Delta Q^{n-p}_W$ relative to Skyrme interactions. Equation
(\ref{PV16}) may also be used to estimate the model spread in $\Delta
Q^{n-p}_W$ for ${}^{133}$Cs. This is given by
%%%
\begin{equation}
\label{PV17}
 \delta \, (\Delta Q^{n-p}_W ) =
      0.9857 N \, \frac{(Z \alpha)^2}{q_p}
      \, \frac{232}{525}
%      \left(\frac{\delta t}{\langle{r_p}\rangle}\right) 
  \, \delta \! \left(\frac{t}{r_p}\right)
      \approx 0.064 \;,
\end{equation} 
%%%
where the model spread for $t/r_p$ has been derived from the $r_n/r_p$
values calculated with $\Lambda_{\rm v}\!=\!0$ and $\Lambda_{\rm 
v}\!=\!0.025$ for ${}^{133}$Cs (listed in Table~\ref{Table2}).
A straightforward estimate of the spread computed as the difference of
the values of $\Delta Q^{n-p}_W$ between the $\Lambda_{\rm v}\!=\!0$
and the $\Lambda_{\rm v}\!=\!0.025$ parameter sets yields
%%%
\begin{equation}
\label{PV18}
 \delta \, (\Delta Q^{n-p}_W ) \approx 0.342-0.278= 0.064 \;,
\end{equation} 
%%%
which compares well with our previous value. In Ref.~\cite{hor4} it
has been argued that for the purpose of significantly reducing the
nuclear structure uncertainties, relative to those arising from
radiative corrections, the neutron radius of ${}^{133}$Cs must be
known within 2\%. The promise of a 1\% measurement of the neutron
radius of $^{208}$Pb at the Jefferson Laboratory~\cite{mic1}, together
with theoretical correlations of the type displayed in Fig.~\ref{Fig4}, 
will be of considerable help in reaching the desired goal. 

% >>>>>>>>>>>>>>>>>>>>>>>>>>>>>>>>>>>>>>>>>>>>>>>>>>>>>>>>>>>>>>>>>>>>
% FRANCIUM

\subsection{A glimpse into the future: Fr isotopes}

Another strategy proposed as a means for attempting to improve the
accuracy in APNC experiments is the utilization of high-$Z$ atoms that
enhance parity violating effects owing to both the increase in the
number of nucleons and the increased electronic density in the
neighborhood the nucleus. Further, alkali atoms have the simplest
atomic structure, thereby minimizing theoretical uncertainties. This
makes francium, the heaviest ``simple'' atom, an attractive candidate
for APNC experiments. Indeed, the increased number of nucleons and the
increased electronic density in the region of the nucleus makes the
APNC effect in Fr some 15 to 20 times larger than in
Cs~\cite{spr1,dzu1}. Unfortunately, francium has no stable naturally
occurring isotopes. Although remarkable experimental progress has
already been achieved in the field~\cite{spr1,Sanguinetti03,Aubin04},
the production of intense beams of radioactive Fr isotopes represents
an exciting challenge for the future of the rare-isotope facilities
all over the world.

\begin{figure}
\includegraphics[width=0.85\linewidth, angle=0, clip=true]{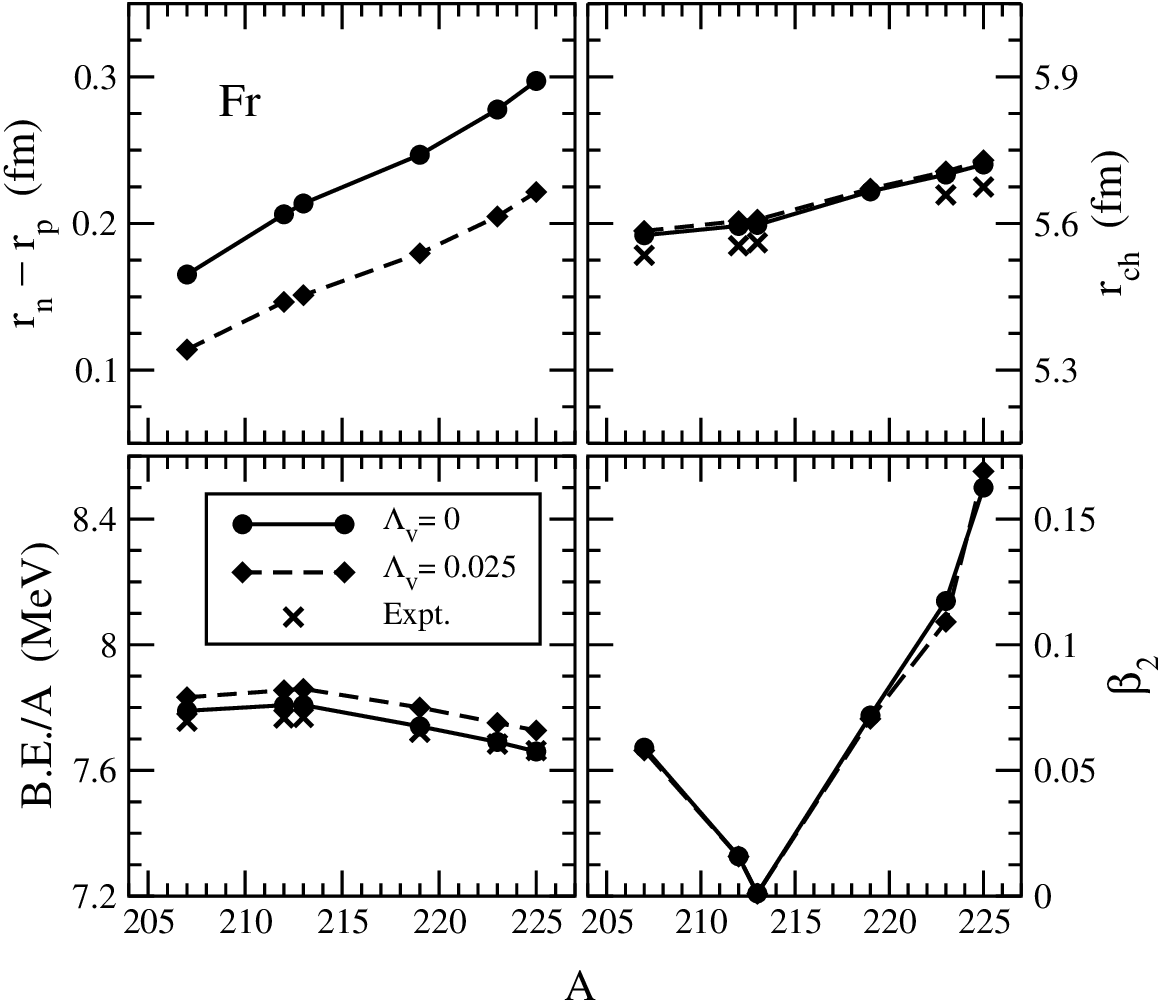}
\caption{\label{Fig5} Calculated ground-state properties (neutron
skin, charge radius, binding energy per particle, and quadrupole
deformation) of some francium isotopes. Experimental data are shown
for comparison when available~\cite{ang1,aud1}.}
\end{figure}

We present estimates for APNC observables in francium along an isotopic 
chain that ranges from $^{207}$Fr to $^{225}$Fr and that includes the two
isotopes with the longest half-life (of about 20 min); these are
$^{212}$Fr and $^{223}$Fr. In Fig.~\ref{Fig5} we display the neutron
skin, charge radii, binding energy per nucleon, and deformation for
six isotopes in the chain computed with the two extremes of the values
of the isoscalar-isovector coupling constant used in this work
($\Lambda_{\rm v}\!=\!0$ and $\Lambda_{\rm v}\!=\!0.025$). As
established earlier for ${}^{174}$Yb in Fig.~\ref{Fig1}, the
calculated values for the charge radii, binding energies, and
deformation show little model dependence and reproduce the existing
experimental data (where available)~\cite{ang1,aud1} rather well. As
expected, it is only the (experimentally unknown) neutron skins that
are sensitive to the value of $\Lambda_{\rm v}$. Note that with the
exception of the semi-magic nucleus $^{213}$Fr, the ground state of
all other isotopes is found to be deformed.

As was done for the Cs isotopes in Table~\ref{Table2}, we display in
Table~\ref{Table3} the calculated weak charge for the six considered Fr
isotopes, alongside other relevant observables. According to
Eq.~(\ref{QWSMRad}), the standard model predictions for the weak charges
of $^{212}$Fr and $^{223}$Fr result in $Q^{\rm SM}_W=-117.380\pm
0.20$ and $-128.223\pm 0.20$, respectively, where the uncertainty in
$Q^{\rm SM}_W$ has been estimated in a similar way to that of $^{133}$Cs
in Ref.~\cite{mar1}. The nuclear corrections to
the weak charge of $^{212}$Fr ($^{223}$Fr) calculated from $q_n$ and
$q_p$ through Eq.~(\ref{PV15}) are given as follows:
\begin{equation}
 \label{PV19}
 \Delta Q^{n-p}_W = 
 \begin{cases}
   1.171 \; (1.584), & \text{if $\Lambda_{\rm v}\!=\!0$,}    \\
   0.905 \; (1.237), & \text{if $\Lambda_{\rm v}\!=\!0.025$.}
 \end{cases}
\end{equation}
These results suggest that nuclear-structure uncertainties in our model,
arising from differences in the shape of the neutron density relative
to that of the proton, are 0.266 and 0.347 for the francium isotopes
$^{212}$Fr and $^{223}$Fr, respectively. The numbers are similar to
the uncertainties 0.264 and 0.345 that one would obtain with the
sharp-radius approximation of Eqs.~(\ref{PV16}) and (\ref{PV17}).

Thus, with the expected precision to be attained by the PREX
experiment, which we mimic here through a change in $\Lambda_{\rm v}$,
our predictions for the weak charge $Q_{W}= Q_W^{\rm SM} + \Delta
Q^{n-p}_{W}$ of the $^{212}$Fr and $^{223}$Fr isotopes vary between
$Q_W=-116.209$ ($\Lambda_{\rm v}=0$) and $-116.475$ ($\Lambda_{\rm
v}=0.025$) for $^{212}$Fr, and between $Q_W=-126.639$ and $-126.986$
for $^{223}$Fr. Note that the uncertainty due to $\Lambda_{\rm v}$
should be augmented by a $\pm0.20$ error from $Q_W^{\rm SM}$.

\section{Conclusions}
\label{Conclusions}
Motivated by the prospects of a high-precision (1\%) measurement of 
the neutron radius of $^{208}$Pb at the Thomas Jefferson Laboratory, 
we have examined the impact of such experiment on atomic parity 
non-conserving observables. While such a measurement will have far 
reaching consequences in fields as diverse as nuclear structure, 
flavor violations in the strong interactions, and nuclear astrophysics, 
we have focused in this contribution exclusively on APNC observables.

Our theoretical framework is based on the highly successful
relativistic NL3 parameter set, suitably modified by the addition of
an isoscalar-isovector term ($\Lambda_{\rm v}$). The virtue of such a
term is that it enables one to change the neutron radius of heavy
nuclei---which at present is poorly known---without compromising the
success of the model in reproducing a variety of ground state
properties of nuclei throughout the periodic table. Relative to an
earlier study based closely on this formalism, the present study
improves on it in one essential aspect: the inclusion of deformation
and pairing correlations. Without adding these effects, the
predictions for the structure of most nuclei along the isotopic chains
of relevance to the APNC program would be unreliable. As a result, in
the present work we have mapped the neutron radii of the relevant
isotopic chains as a function of both $N$ and $\Lambda_{\rm v}$. As
established by others before us, we have found a strong correlation
between the neutron radius of ${}^{208}$Pb and the neutron radius of
APNC nuclei---even for the case of odd-even and odd-odd
nuclei. Employing this correlation, we have used a range of values for
$\Lambda_{\rm v}$ that closely matches the 1\% (or $\sim 0.056$~fm)
uncertainty expected to be achieved by the Jefferson Laboratory
experiment.

With this information at hand, we proceeded to study the impact of
such a 1\% measurement on two different combinations of APNC
observables. The first set of observables (${\cal R}_1$ and ${\cal
R}_2$) are formed from {\it ratios} of weak nuclear charges along
isotopic chains. The merit of such observables is that the ratios are
largely insensitive to uncertainties in atomic structure, leaving
nuclear uncertainties---in the form of the neutron skin---as the main
source of theoretical error. The second set of observables involves a
direct determination of the weak charge of various alkali metals, such
as cesium and francium. Unfortunately, these elements have very few
stable or long-lived isotopes, so in their case the accurate
determination of ratios of weak nuclear charges is more difficult.

In the case of the isotopic ratio ${\cal R}_1$, it has been claimed
that a significant test of the standard model requires a determination
of ${\cal R}_1$ to better than 0.1\%. This precision would be required
for ${\cal R}_1$ to supersede the ${}^{133}$Cs experiment as the most
stringent test for new physics within the APNC program. Unfortunately,
our results indicate that the projected 1\% accuracy in the
measurement of the neutron radius of ${}^{208}$Pb at the Jefferson
Laboratory appears unlikely to translate into the required 0.1\% (or
lower) uncertainty in ${\cal R}_1$. Instead, we have established an
uncertainty in ${\cal R}_1$ that is two to three times larger (of the
order of 0.25--0.35\%). Although the Jefferson Laboratory experiment
is unlikely to achieve the desired accuracy, it is plausible that
second-generation experiments may reach this goal.

In the case of the weak nuclear charge of the alkali metals, where few
stable or long-lived isotopes exist, one must consider theoretical
uncertainties arising from both atomic and nuclear structure. Although
it is no longer possible to fully eliminate uncertainties in atomic
structure (as was done for ${\cal R}_1$) alkali metals at least enjoy
the simplest atomic structure. As far as the nuclear structure is
concerned, the uncertainty is fully subsumed into a single factor: the
neutron skin of the nucleus of interest. It has been argued that to
reduce significantly the nuclear-structure uncertainties relative to
radiative corrections, the neutron radius of ${}^{133}$Cs must be
known with an accuracy of 2\% or better. Similar accuracy should be
expected for the case of other alkali metals. Indeed, in anticipation
of the commissioning of rare-isotope accelerators all over the world
and the ongoing advances in the field of production and
magneto-optical trapping of radioactive atoms, we have computed the
weak nuclear charge of the two longest lived francium isotopes:
$^{212}$Fr (20 min) and $^{223}$Fr (21.8 min). Our assumed uncertainty
in the neutron radii of $^{212}$Fr and $^{223}$Fr translated into a
theoretical uncertainty in the value of their weak nuclear charges of
0.2\% and 0.3\%, respectively.

In summary, we have studied the impact of a high-precision measurement
of the neutron radius of ${}^{208}$Pb on atomic parity
non-conservation experiments. However, the relevance of such a
measurement on a plethora of other research areas---such as nuclear
structure, flavor violations in QCD, and neutron-star structure---has
been strongly emphasized here and elsewhere. Thus, the neutron radius
of ${}^{208}$Pb stands as one of the most fundamental nuclear-physics
quantities yet to be accurately determined. This unfortunate situation
should be promptly corrected.

%%%%%%%%%%%%%%%%%%%

\acknowledgments{This work was partially supported by Grants
BFM2002-01868 (DGI, Spain, and FEDER), 2001SGR-00064 (DGR, Catalonia),
and DE-FG05-92ER40750 (United States Department of Energy). T. S. also
thanks the Spanish Education Ministry grant SB2000-0411 for financial 
support. J. P. thanks the Departament d'Estructura i Constituents de 
la Mat\`eria at the Universitat de Barcelona for its hospitality
during the time that this project was developed.}

% >>>>>>>>>>>>>>>>>>>>>>>>>>>>>>>>>>>>>>>>>>>>>>>>>>>>>>>>>>>>>>>>>>>>
% REFERENCES.
%
%\begin{references}

%\end{references}
%________________________
\newpage
\begin{table}
\caption{The relative uncertainty $\delta {\cal R}_1/{\cal R}_1$ in
the APNC isotope ratio ${\cal R}_1$ and the various components needed
to evaluate it, as described in the text. The last row denotes the
model spread $\delta (\Delta\bar{t})$, Eq.\
(\ref{ModelSpread}), that would be required to achieve $\delta {\cal
R}_1/{\cal R}_1 = 0.1\%$.}
\begin{ruledtabular}
\begin{tabular}{ccccc}
 Observable &  Cs & Ba & Dy & Yb  \\ 
 \hline
 $Z$ & 55 & 56 & 66 & 70   \\
 $N^\prime$ & 82 & 82 & 98 & 106     \\
 $\Delta N$ & 14 &  8 &  8 &   8     \\
 $\langle r_p\rangle$ & 4.752 & 4.766 & 5.134 & 5.254  \\
 \hline
 $\delta(\Delta\bar{t})$ & 0.0059 & 0.0033 & 0.0022 & 0.0022 \\
 $\delta {\cal R}_1/{\cal R}_1$   
                         & 0.0025 & 0.0025 & 0.0027 & 0.0035  \\
 $[\delta(\Delta\bar{t})]_{0.1\%}$
                         & 0.0024 & 0.0013 & 0.0008 & \;0.0007
 \label{Table1}
\end{tabular}
\end{ruledtabular}
\end{table}

%_______________

\begin{table}
\caption{A variety of quantities of relevance to APNC (as
         defined in the text) for four Cs isotopes ($Z=55$), including
         the sole stable element ${}^{133}$Cs. The experimental
         value for the weak charge of this isotope is given by
         $Q^{\rm exp}_W=-72.71(29)_{\rm exp}(39)_{\rm theor}$~\cite{kuc1} and
         $Q^{\rm exp}_W=-72.90(28)_{\rm exp}(35)_{\rm theor}$~\cite{mil1}.}
\begin{ruledtabular}
\begin{tabular}{c|cccccccccc}
$\Lambda_{\rm v}$ & $N$ & $Q^{\rm SM}_W$ & $Q^{\rm nucl}_W$ & $q_n$ &
$q_p$ & $r_n/r_p$ & $r_p$ & $\Delta Q^{n-p}_W$ & $Q_{W}$ \\ 
\hline
0.000 & 76& $-71.226$& 3.284& 0.95410
      & 0.95803& 1.04103& 4.74141& 0.308& $-70.918$\\
      & 78& $-73.197$& 3.400& 0.95377
      & 0.95803& 1.04568& 4.74551& 0.342& $-72.855$\\
      & 80& $-75.169$& 3.512& 0.95350
      & 0.95803& 1.04949& 4.74979& 0.373& $-74.796$\\
      & 82& $-77.140$& 3.622& 0.95327
      & 0.95804& 1.05277& 4.75903& 0.402& $-76.738$\\  
\hline
0.025 &  76& $-71.226$& 3.228& 0.95484
      & 0.95803& 1.03062& 4.75244& 0.249& $-70.977$\\
      &  78& $-73.197$& 3.338& 0.95457
      & 0.95804& 1.03449& 4.75720& 0.278& $-72.919$\\
      &  80& $-75.169$& 3.444& 0.95436
      & 0.95804& 1.03742& 4.76113& 0.302& $-74.867$\\
      &  82& $-77.140$& 3.549& 0.95418
      & 0.95804& 1.04007& 4.76972& 0.325& $-76.815$
 \label{Table2}
\end{tabular}
\end{ruledtabular}
\end{table}

%___________________

\begin{table}
\caption{A variety of quantities of relevance to APNC (as
         defined in the text) for six Fr isotopes ($Z=87$). Francium
         has no stable isotopes.} 
\begin{ruledtabular}
\begin{tabular}{c|ccccccccccc}
$\Lambda_{\rm v}$ & $N$ & $Q^{\rm SM}_W$ & $Q^{\rm nucl}_W$ & $q_n$ &
$q_p$ & $r_n/r_p$ & $r_p$ & $\Delta Q^{n-p}_W$ & $Q_{W}$ \\ 
\hline
0.000 & 120& $-112.451$& 12.650& 0.88788& 0.89510
      & 1.02993& 5.51834& 0.954& $-111.497$\\
      & 125& $-117.380$& 13.361& 0.88659& 0.89510
      & 1.03724& 5.53761& 1.171& $-116.209$\\
      & 126& $-118.366$& 13.502& 0.88636& 0.89510
      & 1.03857& 5.53974& 1.213& $-117.153$\\
      & 132& $-124.280$& 14.294& 0.88544& 0.89511
      & 1.04402& 5.60901& 1.406& $-122.874$\\
      & 136& $-128.223$& 14.866& 0.88454& 0.89511
      & 1.04921& 5.64364& 1.584& $-126.639$\\
      & 138& $-130.194$& 15.171& 0.88397& 0.89512
      & 1.05246& 5.66462& 1.693& $-128.501$\\  
\hline
0.025 & 120& $-112.451$& 12.453& 0.88955& 0.89510
      & 1.02062& 5.52739& 0.734& $-111.717$\\
      & 125& $-117.380$& 13.123& 0.88853& 0.89510
      & 1.02641& 5.54760& 0.905& $-116.475$\\
      & 126& $-118.366$& 13.250& 0.88839& 0.89510
      & 1.02723& 5.55034& 0.932& $-117.434$\\
      & 132& $-124.280$& 14.015& 0.88759& 0.89511
      & 1.03199& 5.61457& 1.094& $-123.186$\\
      & 136& $-128.223$& 14.556& 0.88686& 0.89511
      & 1.03624& 5.65014& 1.237& $-126.986$\\
      & 138& $-130.194$& 14.844& 0.88637& 0.89512
      & 1.03904& 5.67325& 1.329& $-128.865$
 \label{Table3}
\end{tabular}
\end{ruledtabular}
\end{table}

%___________________

\newpage
\end{document}